%% file: main.tex
\documentclass[runningheads]{llncs}

\input{macros}
\usepackage{syntax}
\usepackage[T1]{fontenc} 
\usepackage{textcomp} 
\usepackage{courier} 
\usepackage{listings}
\usepackage[utf8]{inputenc}
\usepackage{booktabs}
\usepackage[table]{xcolor}
\usepackage{graphicx, multirow, booktabs, color, listings}
\usepackage{tikz}
\usetikzlibrary{positioning,calc,fit,backgrounds}
\usepackage{subfig}
\usepackage{makecell}

\begin{document}
\title{Coverage Analysis of Net Inscriptions in Coloured Petri Net Models\thanks{This work was partially supported by the European Horizon 2020 project COEMS under grant agreement no. 732016 (https://www.coems.eu/). We thank Svetlana Jak\v{s}i\'{c} for discussions on this work.}}

\author{Faustin Ahishakiye \and
Jos{\'e} Ignacio Requeno Jarabo \and\\
Lars Michael Kristensen \and
Volker Stolz}
\authorrunning{F. Ahishakiye et al.}

\institute{Dept. of Computer Science, Electrical Engineering, and Mathematical Sciences\\
Faculty of Engineering and Natural Science\\
Western Norway University of Applied Sciences\\
\email{\{fahi,jirj,lmkr,vsto\}@hvl.no}}
\maketitle

\input{abstract}

\section{Introduction}
\input{intro}
\section{Coverage Analysis and MC/DC}
\input{background}

\input{instrumentation}
\input{postprocessing}
\section{Evaluation on Example Models}
\input{evaluation}
\section{Related Work}
\input{related}
\section{Summary and Outlook}
\input{summary}

\bibliographystyle{splncs04}
\bibliography{references}

\end{document}

%% file: macros.tex
\newcommand\CPNTools{CPN Tools}

\newcommand\code[1]{{\ttfamily #1}}

\usepackage[draft]{fixme}

\FXRegisterAuthor{vs}{evs}{VS$\!\!$} 
\FXRegisterAuthor{vp}{VP}{\color{violet}VP$\!\!$}
\FXRegisterAuthor{sj}{SJ}{\color{orange}SJ$\!\!$}

\fxsetup{targetlayout=color}
\fxsetface{margin}{\tiny}
\fxsetup{envface=\tiny\it}  
\fxsetup{inlineface=\tiny\bfseries}

%% file: abstract.tex
\begin{abstract}
High-level Petri net such as Coloured Petri Nets (CPNs) are characterised by the combination of Petri nets and a high-level programming language. In the context of CPNs and CPN Tools, the inscriptions (e.g., arc expressions and guards) are specified using Standard ML (SML). The application of simulation and state space exploration (SSE) for validating CPN models traditionally focusses on behavioural properties related to net structure, i.e., places and transitions.  This means that the net inscriptions are only implicitly validated, and the extent to which these have been covered is not made explicit. The contribution of this paper is an approach that establishes a link between coverage analysis known from programming languages and net inscriptions of CPN models. Specifically, we consider Modified Condition/Decision Coverage (MC/DC) which generalises branch coverage of SML decisions.  We have implemented our approach in a library for CPN Tools comprised of an annotation and instrumentation mechanism that transparently intercept and collect evaluations of boolean conditions, and a post-processing tool that determines whether each decision is MC/DC-covered by a set of models executions (runs).
We evaluate our approach on four larger public-available CPN models.

\end{abstract}


%% file: intro.tex
\label{sec:intro}

Coverage analysis is important for programs in relation to fault detection. Structural coverage criteria are required for software safety and quality design assurance~\cite{KellyJ.:2001:PTM:886632}, and a low coverage indicates that the software product has not been extensively tested. As an example, two common metrics are statement- and branch coverage \cite{10.1145/267580.267590}, where a low coverage indicates that certain instructions have never actually been executed.
Coloured Petri Nets~\cite{CPN2015} and CPN Tools~\cite{CPN2010,cpn2001} have been widely used for constructing models of concurrent systems with simulation and state space exploration being the two main techniques for dynamic analysis. CPN model analysis is generally concerned with behavioural properties related to boundedness, reachability, liveness, and fairness properties. This means that the main focus is on structural elements such as places, tokens, markings (states), transitions and transition bindings. Arc expressions and guards are only implicitly considered via the evaluation of these net inscriptions taking place as part of the computation of transition enabling and occurrence during model execution. This means that design errors in net inscriptions may not be detected as we do not obtain explicit information on for instance whether both branches of an if-then-else expression on an arc have been covered.

We argue that from a software engineering perspective, it is important to be explicitly concerned with quantitative and qualitative analysis of the extent to which net inscriptions have been covered. Our hypothesis is that the coverage criteria used for traditional source code can also be applied to the net inscriptions of CPN models. Specifically, we consider the modified condition decision coverage (MC/DC) criterion. MC/DC is a well-established coverage criteria for safety-critical systems, and is required by certification standards, such as the DO-178C~\cite{pothon2012178c} in the domain of avionic software systems. In the context of MC/DC, a \textit{decision} is a boolean expression composed of sub-expressions and boolean connectives (such as logical ``and''). A \textit{condition} is an atomic (boolean) expression. According to the definition of MC/DC~\cite{chilenski1994applicability,rierson2013developing}, each condition in a decision has to show an independent effect on that decision’s outcome by: (1) varying just that condition while holding fixed all other possible conditions; or (2) varying just that condition while holding fixed all other possible conditions that could affect the outcome. MC/DC is a coverage criterion at the condition level and is recommended due to its advantages of being sensitive to code structure, requiring few test cases ($n+1$ for $n$ conditions), and it is the only criterion that considers the independence effect of each condition.

Coverage analysis for software is usually provided through dedicated instrumentation of the software under test, either by the compiler, or additional tooling, such as binary instrumentation. Transferring this to a CPN model under test, our aim is to combine the execution of a CPN model (by simulation or state space exploration) with coverage analysis of SML guard and arc expressions. In our setting, there is no coverage analysis of the SML expressions integrated into a CPN model. This means that to record coverage data for a CPN model under test, it is necessary to instrument the boolean expressions such that the truth-values of individual conditions are logged in addition to the overall outcome of the decision. Our approach to instrumentation makes use of side-effects by outputting intermediate results of conditions and decisions, which we then process to obtain the coverage verdict. No modifications to the net structure of the CPN model are necessary. Furthermore, we have decoupled the analysis of the coverage data from the model execution so that it does not delay the simulation and state space exploration, and it could even run in parallel.

The rest of this paper is organised as follows. In Section~\ref{sec:background}, we introduce the MC/DC coverage criterion in more detail. In Section~\ref{sec:instrumentation}, we present our approach to deriving coverage data and show how to instrument guard and arc expressions to collect the required coverage data. In Section~\ref{sec:postprocessing} we consider the post-processing of coverage data. We demonstrate the application of our library for coverage analysis on public available 3rd-party CPN models in Section~\ref{sec:evaluation}. In this section, we also evaluate our approach with respect to the currently manual effort for instrumentation, overhead in execution, and discuss our findings.
Section~\ref{sec:related} discusses related work, and we present our conclusions including directions for future work in Section~\ref{sec:summary}. Our coverage analysis library, the example models, the Python code to produce reports and graphs, and documentation is available at \url{https://github.com/selabhvl/cpnmcdctesting}.


%% file: background.tex
\label{sec:background}

There are two main measures of test coverage: requirements coverage and structure coverage~\cite{KellyJ.:2001:PTM:886632}. \textit{Requirements coverage} considers how well requirement- and specification-based test cases verified the implementation, and establishes a relationship between requirements and test cases. \textit{Structural coverage} determines how much of the program/code structure was executed by the requirements-based test cases, and establishes traceability between the code structure and the test cases. Normally, requirements coverage analysis precedes structural coverage analysis. However, requirements may not have a complete specification of all behaviours present in the executable code. In addition, requirements may not be specified at a sufficient level of granularity to assure full testing of all functional behaviour of the code. Hence, requirements-based testing alone cannot confirm that the code does not include bugs.

When considering CPN models under test, we do not generally know the requirements underlying the construction of the model. Furthermore, we cannot assume the explicit presence of test cases as they will only be given implicitly via the behaviour of the model and its initial marking (state). In the terminology of coverage analysis of code, we will therefore be concerned with structural (code) coverage analysis of guard- and arc expressions. For these expressions, the test cases will arise as the transition bindings (occurrence modes) in which these expressions happens to be evaluated during a simulation or a state space exploration of the model.

A guard expression is a list of Boolean expressions all of which are required to evaluate to true in a given transition binding for the transition to be enabled. We refer to such Boolean expressions as \textit{decisions}. Similarly, an if-then-else expression on an arc will have a decision determining whether the then- or the else-branch will be taken. Decision are in turn constructed from \textit{conditions} and Boolean operators according to the definitions below.

\begin{definition}[Condition]
	A \textbf{condition} is a Boolean expression containing no Boolean operators except for the unary operator NOT.
\end{definition}

\begin{definition}[Decision]
	A \textbf{decision} is a Boolean expression composed of conditions and zero or more Boolean operators. It is denoted by  D($c_1,c_2,c_i,\cdots, c_n$), where $c_i$, $1 \leq i \leq n$ are Boolean conditions. 
\end{definition}

As an example, we may have a guard (or an arc expression) containing a decision of the form $D$ = $(a \wedge b) \vee c$, where $a$, $b$, and $c$ are conditions. These conditions may in turn refer to the values bound to the variables of the transition.

The evaluation of a decision requires a \textit{test case} assigning a value to the conditions of the decision according to the following definition.

\begin{definition}[Test case]
	Given a decision $D$, a \textbf{test case} is a truth vector (also called a test-evaluation) $TC=(e_1, e_2, e_3,\cdots, e_n)$ where $e_i \in \{0,1,?\}$ is the value assignment to condition $c_i$ of $D$, with $?$ meaning that a condition was not evaluated due to short-circuiting (see below). A \textbf{test suite} for a decision is a set of test cases $\{TC_1,TC_2,\cdots,TC_k\}$.
\end{definition}

Different software levels (A-D) require different structure coverage criteria,  \textit{statement} coverage for software levels A-C, \textit{branch/decision} coverage for software levels A-B, and MC/DC for software level A~\cite{10.1145/267580.267590}.
 Statement coverage is considered inadequate because it is insensitive to some control structures. That is, if there is no test case that causes a conditional statement to evaluate false, statement coverage rates the code fully covered, but the code may fail, if a condition ever evaluates false~\cite{Steve}. In addition, it does not report whether loops reach their termination condition, only whether the loop body was executed. Both statement- and branch coverage are completely insensitive to the logical operators ($\vee$ and $\wedge$). In ~\cite{KellyJ.:2001:PTM:886632}, coverage criteria taking logical expressions into consideration have been defined and proposed. These are \textit{condition coverage} (CC), where each condition in a decision takes on each possible outcome at least once true and at least once false during testing; \textit{decision coverage} (DC) requiring only each decision to be evaluated once true and once false; and \textit{multiple condition coverage} (MCC) which is an exhaustive testing of all possible input combinations of conditions to a decision. CC and DC are considered inadequate due to ignorance of the independence effect of conditions on the decision outcome. MCC requires $2^n$ tests for a decision with $n$ inputs. This results in exponential growth in the number of test cases, and is therefore time-consuming and impractical for many test cases.

 DC has also the disadvantage that it ignores branches within Boolean expressions which occur due to short-circuit operators. Short-circuit means that the right operand of the \textit{and}-operator (\&\&/$\wedge$)
 is not evaluated if the left operand is false, and the right operand of the \textit{or}-operator (||/$\vee$)
 is not evaluated if the left operand is true.  Consider an example in Listing \ref{ref:Shortcircuit}, the decision is evaluated to true when condition1 and condition2 are true whereas function1 is short-circuited in that case. When condition1 is false, the decision is evaluated to false, condition2 is not evaluated and there is no call to function1.

 \begin{lstlisting}[basicstyle=\small, caption={Illustration of short-circuit evaluation}\label{ref:Shortcircuit}]
 if (condition1 && (condition2 || function1()))
    statement1;
 else
    statement2;
 \end{lstlisting}

To address the limitations of the structure coverage criteria discussed above, \textit{modified condition/decision coverage} (MC/DC) is considered. In safety critical systems such as in the avionics industry, software certification requires a vendor to demonstrate that the test-suite provides MC/DC coverage of the source code. The MC/DC coverage criterion has been chosen as the coverage criterion for the highest safety level software because it is sensitive to the complexity of the decision structure~\cite{chilenski1994applicability}. Compared to even stronger criteria like multiple condition coverage (MCC), that requires every possible combination of all conditions, MC/DC may be satisfied with only $n+1$ test cases for a decision with $n$ conditions~\cite{KellyJ.:2001:PTM:886632,Chilenski01aninvestigation}. In addition, MC/DC coverage criterion is suggested as a good candidate for model-based development (MBD) using tools such as Simulink and SCADE~\cite{4702848}. Therefore, our model coverage analysis is based on MC/DC as a coverage criterion subsuming the other coverage criteria.  The following MC/DC coverage definition is based on DO-178C~\cite{rierson2013developing}:

\noindent\begin{definition}[Modified condition/decision coverage]
	 A program is \\ MC/DC covered and satisfies the MD/DC criterion if the following holds:

	\begin{itemize}
		\item every point of entry and exit in the program has been invoked at least once,
		\item every condition in a decision in the program has taken all possible outcomes at least once,
		\item every decision in the program has taken all possible outcomes at least once,
		\item each condition in a decision has shown to independently affect that decision’s outcome by:
		(1) varying just that condition while holding fixed all other possible conditions, or
		(2) varying just that condition while holding fixed all other possible conditions that could affect the outcome.
	\end{itemize}
\end{definition}

 To demonstrate MC/DC, a structural coverage analysis tool should monitor statements, entry and exit
points, decision and branching statements, and Boolean conditions~\cite{KellyJ.:2001:PTM:886632}.
However, the first item in the definition of MC/DC, is traditionally added to all control-flow criteria and is not directly connected  with the main point of MC/DC~\cite{rcdc}. The most challenging and most discussed part is showing the independent effect. Therefore in our analysis, we are interesting in evaluation of expressions by checking the independence effect of each condition.
\begin{example}
	\label{ex:example1}
	Consider the decision $D$ = $(a \wedge b) \vee c$. The truth table representing all eight possible test cases (combinations) for MCC is given in Table~\ref{tb:MCDCpairs}. In the table, the MC/DC column lists conditions (here $a$,$b$, and $c$) together with a pair of test cases that demonstrate the independence effect of the particular condition. For an example, the MC/DC pair $c(1,2)$ specifies that from test case $1$ and $2$ we can observe that changing the truth value of $c$ while keeping the values of $a$ and $b$, we can affect the outcome of the decision.  Comparing MCC to MC/DC in terms of the number of test cases, there are seven possible MC/DC test cases (test cases 1 through 7) that are part of an MC/DC pair, where condition $c$ is represented by 3 pairs of test cases showing the independence effect of condition $c$, and one pair of test cases for conditions $a$ and $b$. However, all seven test cases provided in Table~\ref{tb:MCDCpairs} are not necessary to ensure MC/DC coverage. Only four test cases (1,2,3, and 4), i.e., $n+1$ test cases for a decision with three conditions are required to achieve MC/DC coverage as shown in Table~\ref{tb:MinMCDCpairs}.
\end{example}
\begin{table}[t]
	\centering
	\subfloat[MCC test cases]{%
		\label{tb:MCDCpairs}
		\scalebox{1.2}{%
		\begin{tabular}{| c | c | c | c | c | c |}
			\hline
			TC & a & b & c & $D$ & MC/DC pairs\\ \hline
			1 & 0 & 0 & 0 & 0                     &             \\ \hline
			2 & 0 & 0 & 1 & 1                     &  c(1,2)		\\ \hline
			3 & 0 & 1 & 0 & 0                     & 	   		\\ \hline
			4 & 0 & 1 & 1 & 1                     &  c(3,4)		\\ \hline
			5 & 1 & 0 & 0 & 0                     &  		\\ \hline
			6 & 1 & 0 & 1 & 1                     &  c(5,6)		\\ \hline
			7 & 1 & 1 & 0 & 1                     &  a(3,7), b(5,7)		\\ \hline
			8 & 1 & 1 & 1 & 1	                  &                 	\\ \hline
		\end{tabular}}%
	}%
	\qquad
	\subfloat[MC/DC test cases]{%
		\label{tb:MinMCDCpairs}
			\scalebox{1.2}{%
		\begin{tabular}{| c  | c | c | c | c | c |}
			\hline
			TC & a & b & c & $D$ & MC/DC pairs\\ \hline
			1 & 0 & ? & 0 & 0                     & 	   		\\ \hline
			2 & 1 & 1 & ? & 1                     &  a(1,2)		\\ \hline
			3 & 1 & 0 & 0 & 0                     &  b(2,3)		\\ \hline
			4 & 0 & ? & 1 & 1                     &  c(1,4)		\\ \hline
		\end{tabular}
	}%
}%
\vspace*{1em}
	\caption{MCC and MC/DC test cases for decision $D$ = $(a \wedge b) \vee c$ }
	\label{tb:MCDCevaluation}
\end{table}
By showing the independent effect of each condition, MC/DC demonstrates that each condition of the decision has a defined purpose. The most challenging and most discussed part in the definition of MC/DC is showing this independent effect: item (2) in the definition has been introduced in the DO-178C to clarify that so-called \textit{Masked MC/DC} is allowed~\cite{cas2001rationale,pothon2012178c}. Masked MC/DC means that it is sufficient to show the independent effect of a condition by holding fixed only those conditions that could actually influence the outcome. This is important for programming languages that use short-circuit evaluation, because certain executions of decisions are not distinguishable, if the outcome of the decision is determined before every condition has been evaluated.


%% file: instrumentation.tex
\section{Instrumentation of CPN models}
\label{sec:instrumentation}

In this section, we describe our instrumentation approach on an example CPN model, and highlight the salient features of our coverage analysis library. Our overall goal is that through simulation or state space exploration, we (partially) fill a truth-table for each decision in the net inscriptions of the CPN model.  Then, for each of these tables, and hence the decisions they are attached to, we determine whether the model executions that we have seen so far satisfy the MC/DC coverage criteria. If MC/DC is not satisfied, either further simulations are necessary, or if the state space is exhausted, developers need to consider the reason for this short-coming, which may be related to insufficient exploration as per a limited set of initial markings considered, or a conceptual problem in that certain conditions indeed cannot contribute to the overall outcome of the decision.

\subsection{MC/DC coverage for CPN models}
MC/DC coverage (or any other type of coverage) is commonly used with executable programs: which decisions and conditions were evaluated by the test cases, and  with which result. Specifically, these are decisions \textit{from to the source code} of the system (application) under test. Of course, a compiler may introduce additional conditionals into the code during code generation, but these are not of concern.

\CPNTools{} already reports a primitive type of coverage as part of simulation (the transition and transition bindings that have been executed) and the state space exploration (transitions that have never occurred). These can be interpreted as variants of state- and branch coverage.

Hence, we first need to address what we want MC/DC coverage to mean in the context of CPN models. If we first consider guard expressions on transitions, then we have two interesting questions related to coverage: if there is a guard, we know from the state space (or simulation) report whether the transition has occurred, and hence whether the guard expression has evaluated to true. However, we do not know if during the calculation of enabling by CPN Tools it ever has been false. If the guard had never evaluated to false, this may indicate a problem in the model or the requirements it came from, since apparently that guard was not actually necessary. Furthermore, if a decision in a guard is a complex expression, then as per MC/DC, we would like to see evidence that each condition contributed to the outcome. Neither case can be deduced from the state space exploration or via the CTL model checker of CPN Tools as the executions only contain transition bindings that have occurred, and hence cases where the guard has evaluated to true.

\subsection{Instrumentation of Net Inscriptions}
\label{instrumentation}

In the following, we describe how we instrument the guards on transitions such that coverage data can be obtained. Arc expressions are handled analogously. Guards in a CPN model are written following the general form of a comma-separated list of boolean expressions (decisions):
\[
  [ bExp_0, \ldots, bExp_n]
\]

\noindent
A special case is the expression
\[
  var = exp
\]

\noindent
which may have two effects: if the variable \code{var} is bound already via a pattern in another expression (arc or guard) of the transition, then this is indeed a boolean equality test (decision). If, however, \code{var} is not bound via other expressions, then this essentially assigns the value of \code{exp} to the variable \code{var} and does not contribute to any guarding effect.

We consider general boolean expressions which may make use of the full feature set of the SML language for expressions, most importantly boolean binary operations, negation, conditional expressions with if-then-else and function calls. Simplified, we handle:

\begin{grammar}
  <bExp> ::= "not" <bExp> | <var> | f <exp>$_0$ \ldots <exp>$_n$\\
  | <bExp> "andalso" <bExp> | <bExp> "orelse" <bExp>\\
  | "if" <bExp> "then" <bExp> "else" <bExp>\\
  | "let" $\ldots$ "in" <bExp> "end"
\end{grammar}

Function symbols \code{f} cover user-defined functions as well as (built-in) relational operators such as \code{$<$,$=$};
we do not detail the overall nature of arbitrary expressions, but refer the reader to \cite{Tofte:2009} for a comprehensive discussion.

State space exploration or simulation of the CPN model is not in itself sufficient to determine the outcome of the overall expression and its subexpressions: guards are not explicitly represented, and we only have the event of taking the transition in the state space, but no value of the guard expressions. Hence, we need to rely on side-effects during model execution to record the intermediate results. Our key idea is to transform every subexpression and the overall decision into a form which will use SML's file input/output to emit a log-entry that we can later collect and analyse. Alternatively, we could have implemented MC/DC coverage analysis in SML, but we chose to reuse an analysis in Python that is more accessible to other potential users.

For the necessary instrumentation which can be viewed as a transformation of guard and arc expressions, we essentially create an interpreter for boolean expressions: when guards are checked (in a deterministic order due to SML's semantics from left to right), we traverse a term representation of the boolean expression and output the intermediate results. Correspondingly, we design a data type (see Listing \ref{ref:datatype}) that can capture the above constructs, and define an evaluation function (see Listing \ref{ref:eval}) on it. As we later need to pinpoint where in a model a problem with coverage occurred, for overall expressions \lstinline{EXPR} and atomic proposition \lstinline{AP} we additionally introduce a component of type string that allows this identification.

\begin{minipage}[t]{0.4\linewidth}
  \begin{lstlisting}[basicstyle=\ttfamily\scriptsize, caption={Expressions}\label{ref:datatype}]
 datatype condition =
      AND of condition * condition
    | OR of condition * condition
    | NOT of condition
    | ITE of condition * condition
                       * condition
    | AP of string * bool;
  \end{lstlisting}
\end{minipage}
\quad
\begin{minipage}[t]{0.45\linewidth}
  \begin{lstlisting}[basicstyle=\ttfamily\scriptsize, caption={Evaluation function}\label{ref:eval}]
fun eval (AP (cond,v))=([(cond, SOME v)],v)
  | eval (OR (a,b)) = let
	val (ares,a') = eval a;
	val (bres,b') = eval b;
    in
	(ares^^bres, a' orelse b')
        end
        ...

fun EXPR (name,expr) : bool = [ ... ]
  \end{lstlisting}
\end{minipage}

The evaluation function \code{eval} collects the result of intermediate evaluations in a list data structure, and the \lstinline{EXPR} function (implementation not shown) turns this result into a single boolean value that is used in the guard, and as a side-effect outputs the truth outcome for individual conditions.

As an example, if we consider a guard:\\

\lstinline{a > 0 andalso (b orelse (c = 42))}\\

\noindent
then we can transform this guard in a straight forward manner into\\

\lstinline{EXPR("Gid", AND(AP("1", a>0), OR(AP("2",b), AP("3", c=42))))}\\

It is important to notice that this does not give us the (symbolic) boolean expressions, as we still leave it to the standard SML semantics to evaluate the \lstinline{a>0}, while abstractly we refer to the AP as a condition named ``1''.
In the following we also elide expression- and proposition names for clarity in the text when not needed.

\begin{figure}
	\centering
	\subfloat[Original model]{\includegraphics[width=.49\linewidth]{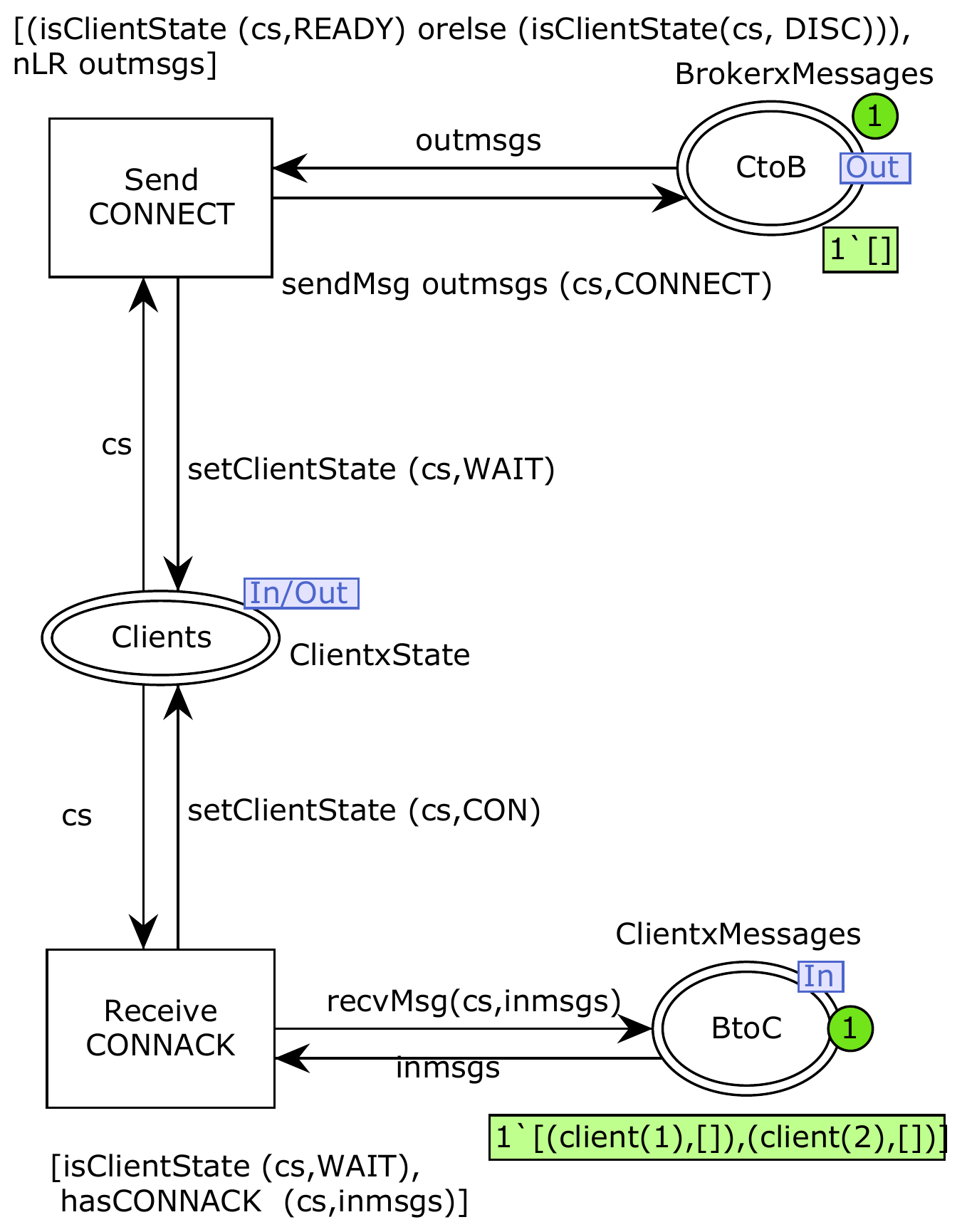}}%
	\hfill
	\subfloat[Instrumented model]{\includegraphics[width=.48\linewidth]{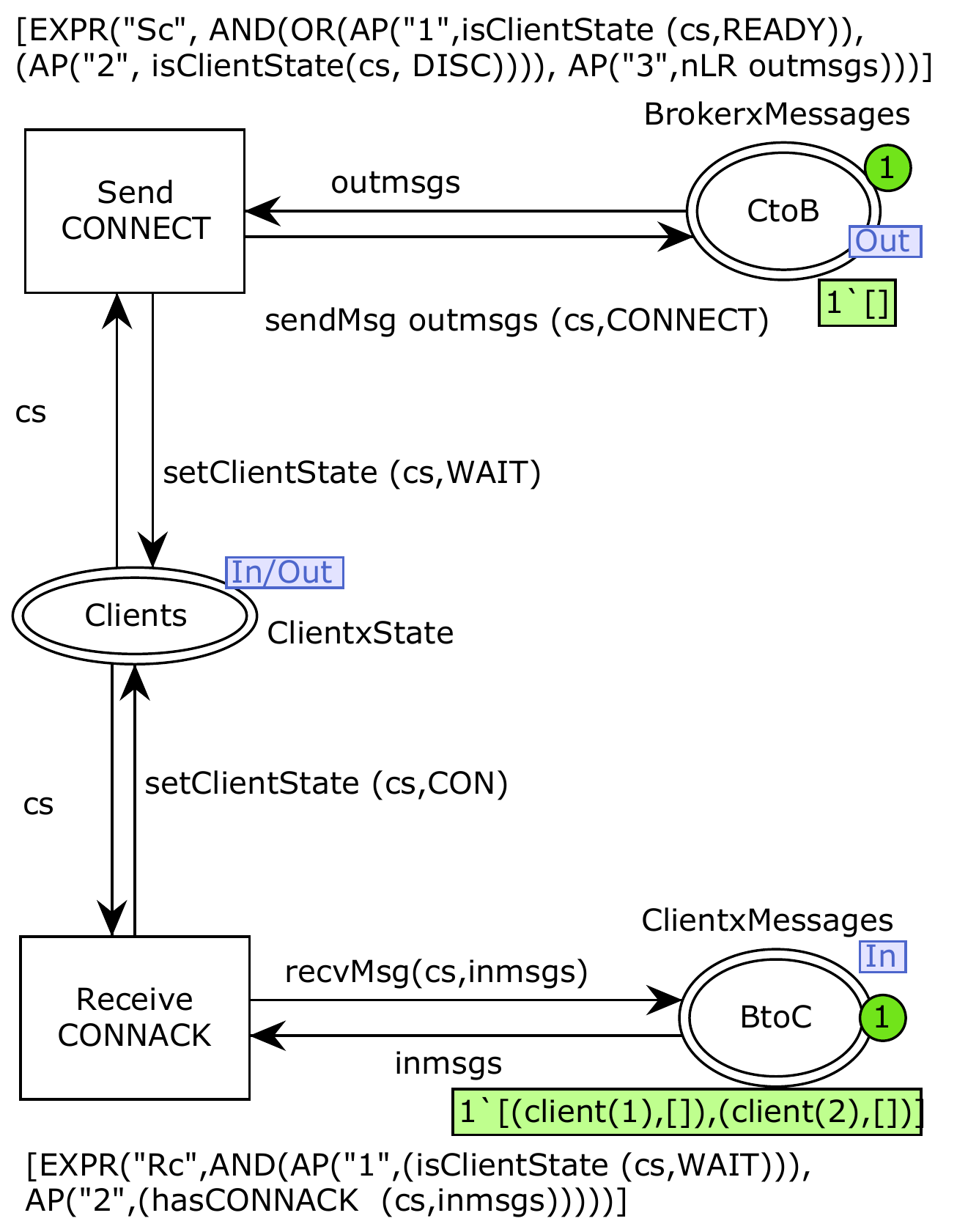}}%
	\caption{MQTT~\cite{Rodrguez2019FormalMA}: guards before and after instrumentation}
	\label{fig:CONNECT}
\end{figure}

Figure~\ref{fig:CONNECT} shows a module from one of our example models \cite{Rodrguez2019FormalMA} before and after instrumentation. The guard expressions for the transitions named \textsf{Send CONNECT} and \textsf{Receive CONNACK} have been instrumented. As the \lstinline{EXPR} function contains input/output statements, we now as a side-effect observe intermediate results of conditions every time a guard is evaluated during execution of the CPN model. Note that the \textsf{CONNECT} module is one of several modules of the MQTT protocol model, and the arc and guard expressions in the other modules were transformed in a similar manner. Even though we have done the instrumentation by hand, it could be automated based on the \code{.cpn} XML file of CPN Tools in combination with an SML parser.

\subsection{Emitting Coverage Data}

\begin{figure}
	\begin{tabular}{cc}
	\includegraphics[width=.5\textwidth]{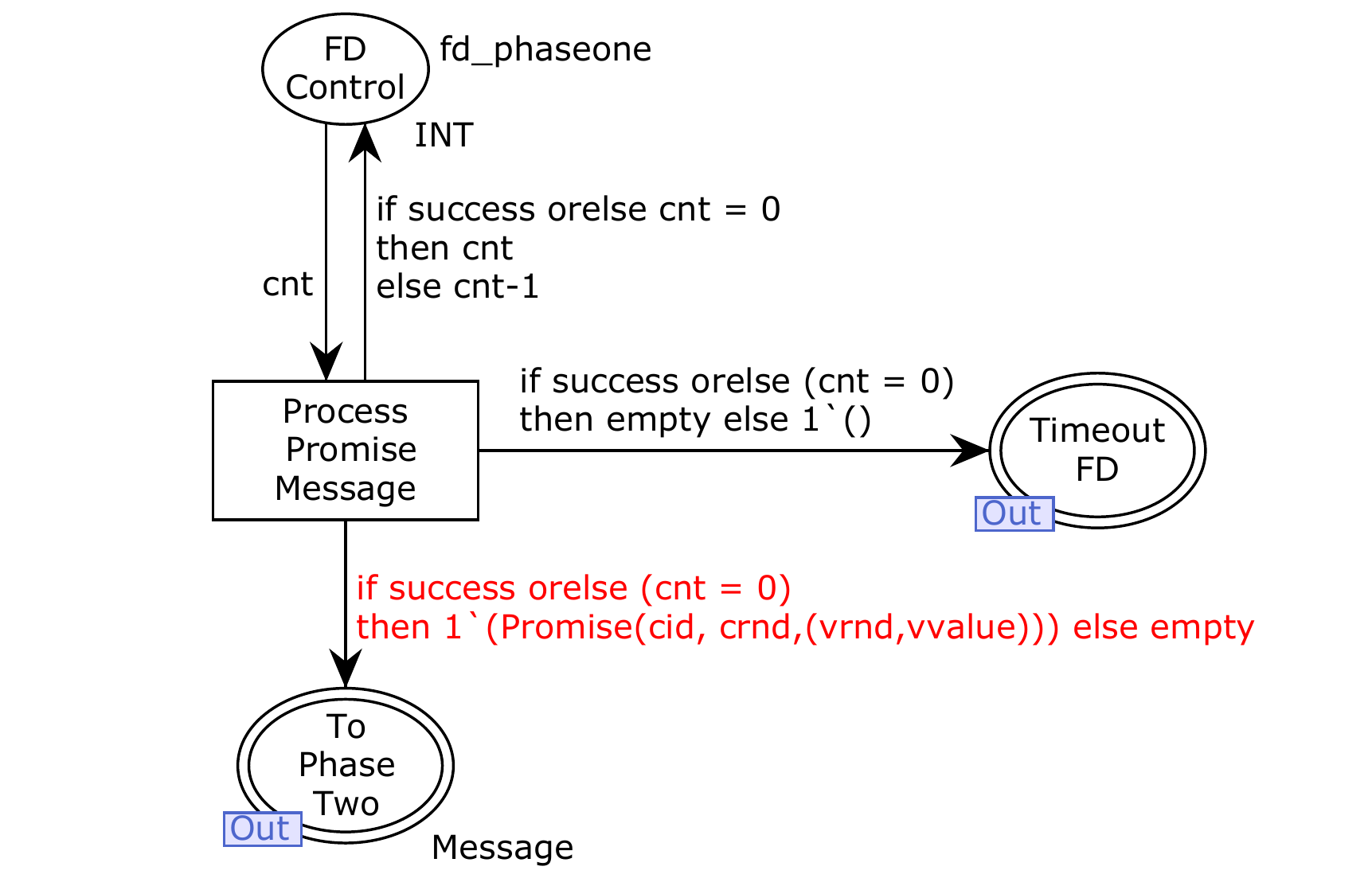}
	&
	\includegraphics[width=.5\textwidth]{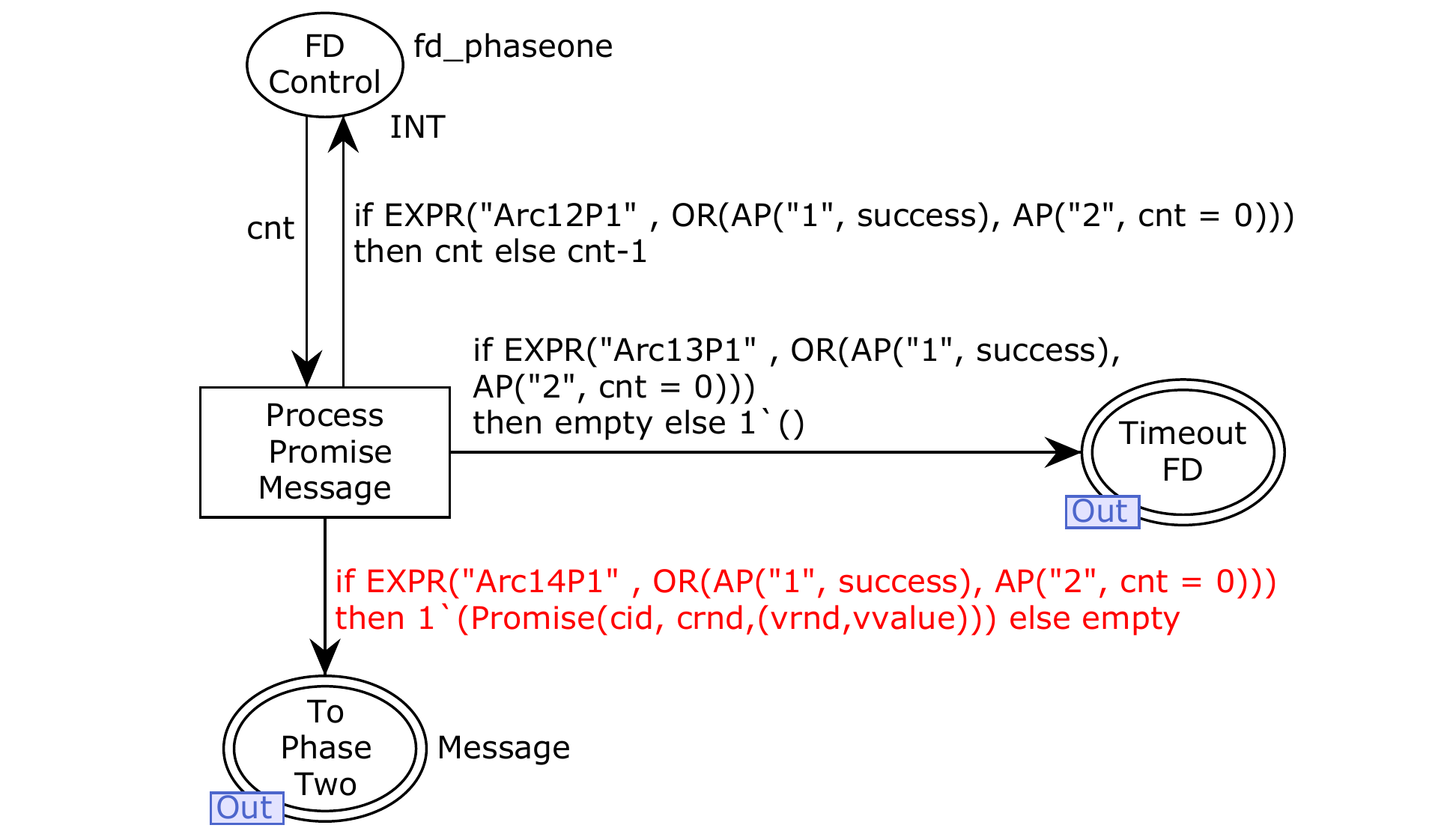}
\end{tabular}
	\caption{Paxos~\cite{WANG2019254}: Arc expression before (left) and after instrumentation (right)}
	\label{fig:arcPhase1}
\end{figure}

 We can likewise apply the transformation to boolean expression in arc expressions:

 {any} boolean expression is transformed into an \lstinline{EXPR($\ldots$(AP bExp)$\ldots$)}, resulting for example in the transformation of \[\mbox{\lstinline{if bexp1 orelse bexp2 then e1 else e2}}\] into \[\mbox{\lstinline{if EXPR(OR(AP bexp1, AP bexp2)) then e1 else e2}.}\] Figure~\ref{fig:arcPhase1} illustrates such a transformed arc expressions for a module of the Paxos protocol which we also use in our evaluation.

Although the observation of conditions would also happen when e.g. using simulation, we rather envisage that this feature is used with state space exploration whenever applicable which in general will evaluate every guard multiple times with varying bindings. The execution of the CPN model is triggered through the standard user interface of \CPNTools{}.

In the scenarios anticipated here in this paper, we assume any subexpressions are \textit{total} and do not crash when evaluated, which would abort the model execution and lead to partial results (up to the crash).
In contrast to SML's \lstinline{orelse}, this also means e.g. for the \lstinline{OR}-construct that we \emph{always} evaluate both arguments, even though \CPNTools{} would make use of short-circuit evaluation. We see the advantage in computing as many outcomes as possible instead of leaving gaps in the truth-tables, as this will give us more precise information to work with when checking the MC/DC criteria.


%% file: postprocessing.tex
\section{Post Processing of Coverage Data}
\label{sec:postprocessing}

We now discuss the coverage analysis which is performed via post-processing of the coverage data recorded through the instrumentation. We did not implement the MC/DC coverage analysis in SML directly. Rather, we feed individual observations about decision outcomes and their constituent conditions into a Python tool that computes the coverage results. This allows us to reuse the backend in other situations, without being SML or CPN specific.

\subsection{Coverage Analysis}
The general format from the instrumentation step is a sequence of colon-delimited rows, where each triple in a row captures a single decision with the truth values of all conditions in a fixed order and the outcome. As an example, see Script 4.1.
The name (stemming from the first argument to an \lstinline{EXPR} above) is configurable and should be unique in the model; we recommend to derive the name from the element (guard or arc) the expression is attached to. This makes it easy to later trace coverage results for this name back to the element in the model, and for the user to navigate to the sub-module containing the element should they desire to do so.\\[1pt]

\begin{minipage}[t]{0.35\linewidth}\textit{Script 4.1: Log decisions}
	{\scriptsize\small
		\begin{verbatim}
		...
		a3:01:0
		t42:01110:0
		t42:01011:1
		...
		\end{verbatim}
	}
\end{minipage}
\begin{minipage}[t]{0.65\linewidth}\textit{Script 4.2: Decisions evaluation table}

	{\ttfamily\scriptsize\small

		\begin{tabular}{ll}
			\ldots\\
			Returna19	&   \\
			0001	&	0 \\
			0010	&	0 \\
			0101	&	0 \\
			0110	&	0 \\
			1001	&	1 \\
			1101	&	1 \\
			1110	&	1 \\
			\ldots\\
		\end{tabular}\\
		
		MCDC covered? False\\
		R\{1:[(0001, 1001), (0101, 1101), (0110, 1110)], 2:[], 3:[], 4:[]\}
	}\\[1pt]
\end{minipage}

Script 4.1 shows that the decision ``$t42$'' was triggered twice, possibly on a guard which did not enable the transition (result column indicating false), after which the exploration choose different transition bindings which resulted in a changed outcome to the 3rd and 5th condition in this decision and an overall outcome of true.

We chose to print the binary representation instead of, e.g., a slightly shorter integer value to facilitate casual reading of the trace.
Also, this allows us to enforce the correct number of bits that we expect per observation, corresponding to the number of conditions in the decision, which mitigates against instrumentation- or naming-mistakes.

Our Python tool parses the log file and calculates coverage information. It prints the percentage of decisions that are MC/DC and branch covered in textual mode and in GNU Plot syntax (see Figures \ref{fig:result} later). The output contains individual reports in the form of the truth tables for each decision, which summarizes the conditions that are fired during the execution of the CPN model, and sets of pairs of test cases per condition that show the independence effect of that condition.

In the case that the decision is not MC/DC covered, the information provided by the Python script helps to infer the remaining valuations of the truth tables that should be evaluated in order to fulfill this criteria.
In the example in Script 4.2, the first condition (left-most column in the table) has multiple complementary entries where the expression only varies in one bit (e.g., rows 0001 and 1001) and the output changes (0 to 1). The \texttt{R} set shows three such pairs for condition 1, but no complementary entries at all are found in the truth table for conditions 2, 3 and 4, and hence indicated as empty sets \code{[]} by Python.
This information can then be used by developers to drill down into parts of their model, e.g.\ through simulation, that have not been covered adequately yet.

\subsection{Combining Coverage Data from Multiple Runs}

Coverage- or testing frameworks rely on their correct use by the operator, only a sub-class of tools such as fuzzers are completely automated.
Our library is ``drag-and-drop'' in the sense that the user only imports it, and invokes the central \code{mcdcgen()} function, in line with how state space exploration works in \CPNTools{}.
This function only explores the state space for the current configuration as determined by the initial markings. Compared to regular testing of software, this corresponds to providing a single input to the system under test.

It is straightforward to capture executions of multiple runs: our API supports passing initialisation functions that reconfigure the net between subsequent runs. However, as there is no standardised way of configuring alternative initial markings or configurations in \CPNTools{}, the user has to actively make use of this API. In the default configuration, only the immediate net given in the model is evaluated.

 \begin{lstlisting}[caption={MC/DC tool invocation}\label{ref:mcdcconfig},  keywords={use, mcdcgen, mcdcgenConfig}]
use (cpnmcdclibpath^"config/simrun.sml");

(* Invocation with default settings (no timeout) *)
mcdcgen("path/to/mqtt.log");

(* Invocation without timeout; base model + 2 configurations *)
mcdcgenConfig(0, applyConfig,[co1,co2],"path/to/mqtt3.log");
\end{lstlisting}

As an example, we show in Listing \ref{ref:mcdcconfig} how we make use of this feature in the MQTT-model, where alternative configurations were easily discoverable for us: the signature of MC/DC-generation with a simple test-driver is
$$\mbox{\lstinline{mcdcgenConfig = fn : int*('a->'b)*'a list*string->unit},}$$
where the first argument is a timeout for the state space exploration, the second is a function with side-effects that manipulates the global configurations that are commonly used in \CPNTools{} to parameterise models, the next argument is a list of different configurations, followed by the filename for writing results to. This function will always first evaluate the initial model configuration, and then have additional runs for every configuration. Internally, it calls into \CPNTools{}' \lstinline{CalculateOccGraph()} function for the actual state space exploration.

Hence the first \lstinline{mcdcgen}-invocation in Listing \ref{ref:mcdcconfig} will execute a full state space exploration without timeout,
whereas the second invocation would produce three subsequent runs logged into the same file, again without a default timeout.
The test-driver can easily be adapted to different scenarios or different ways of reconfiguring a model.
Alternatively, traces can also be produced in separate files that are then simply concatenated for the coverage analysis.


%% file: evaluation.tex
\label{sec:evaluation}

In this section, we provide experimental results from an evaluation of our approach to model coverage for CPNs.
We present the results of examining four non-trivial models from the literature that are freely available as part of scientific publications:
a model of the Paxos distributed-consensus algorithm~\cite{WANG2019254}, a model of the MQTT publish-subscribe protocol~\cite{Rodrguez2019FormalMA}, a model for distributed constraint satisfaction problem (DisCSP) algorithms~\cite{Pascal2017ACP}, and a complex model of the runtime environment of an actor-based model~\cite{GKOLFI20191} (CPNABS).
All models come with initial markings that allow state space generation, in the case of MQTT and DisCSP finite, and infinite---and hence possibly partial---in the case of Paxos and CPNABS.

\subsection{Experimental Setup}

\begin{figure}[bhtp]
	\centering
	\includegraphics[trim=4.5cm 4cm 9cm 1.0cm, clip=true, width=0.65\linewidth]{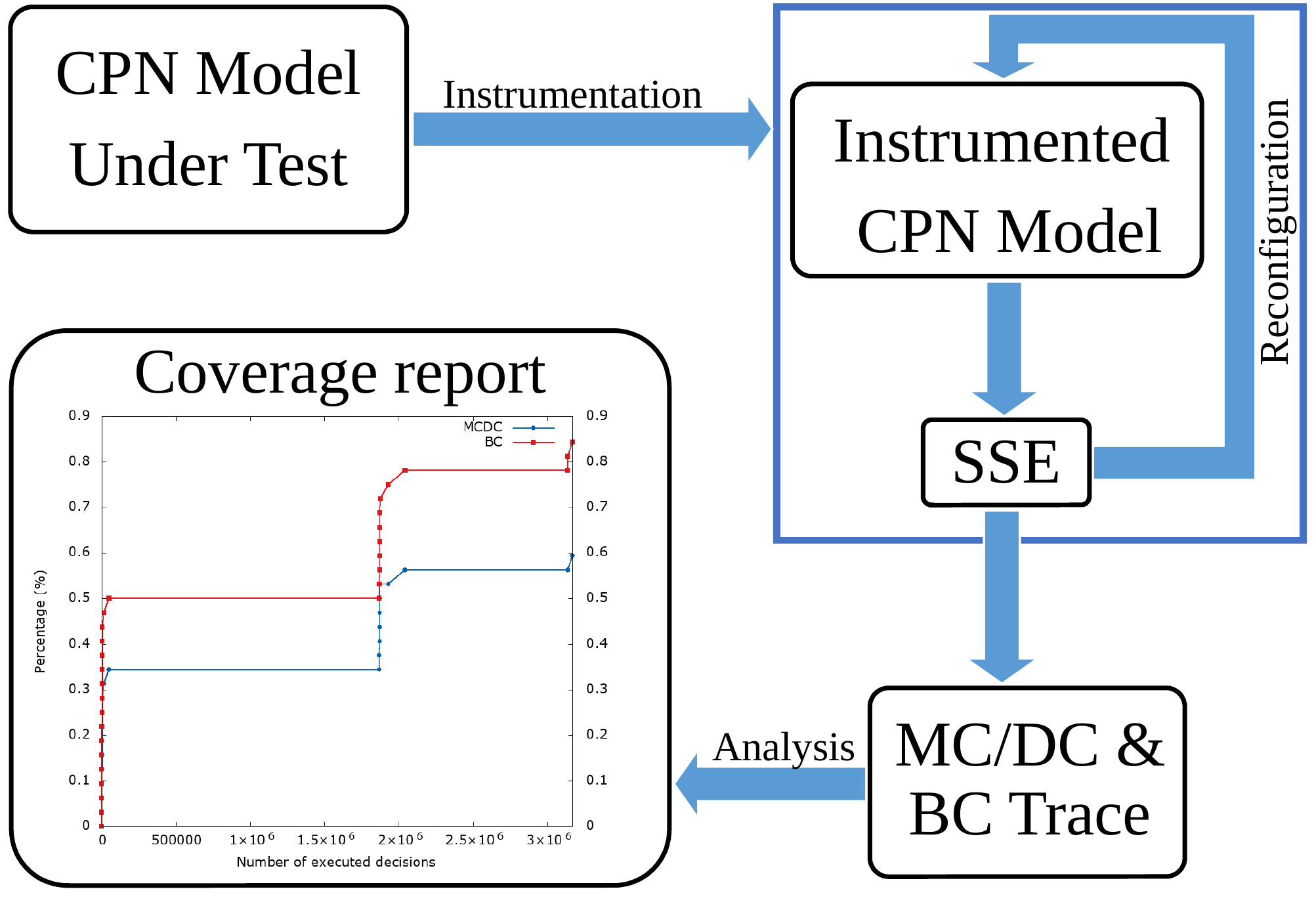}
	\caption{Experimental setup}
	\label{fig:setup}
\end{figure}

Figure~\ref{fig:setup} gives an overview of our experimental setup.
Initially, we have the original CPN model under test. The first step is to instrument the model by transforming each guard and arc expression into a form that as a side-effect prints how conditions were evaluated and the overall outcome of the decision (cf. Section~\ref{sec:instrumentation}). In the second step, we run the state space exploration (SSE) on the instrumented model and then reconfigure the configuration (initial marking). As the side effect of SSE, we run the MC/DC generation which gives as output a log file containing the information of evaluations of conditions in arcs expressions and guards and the decision outcome. The final step is to run the MC/DC analyser which is a post-processing tool that determines whether each decision is MC/DC-covered or not.
The MC/DC analyser is implemented as a Python script that checks the independence effect of each condition on the decision outcome based on the input log file and outputs the MC/DC coverage results.
In addition, it reports the branch coverage (BC), by checking if each of the possible branches in each decision has been taken at least once.

\subsection{Experimental Results}

Table \ref{tb:results} presents the experimental results for the four example models \cite{GKOLFI20191,Rodrguez2019FormalMA,WANG2019254,Pascal2017ACP}.
The CPN model under test contains Boolean expressions in arcs and guards, and the total number of expressions (both guard and arc expressions) is denoted by $m$. For each model, we consider the number of executed decisions (second column) in arcs and guards.
 Column \textit{Model decisions ($m$)} refers to the number of decisions that have been instrumented in the model.
The number of decisions observed in the model and in the log-file may deviate in case some of decisions are never executed in which case they will not appear in the log file.
We indicate in brackets if during our exploration we did not visit, and hence log, each decision at least once;
in the case of DisCSP, there are two guards decisions which never executed.
 Since the main concern of MC/DC is the independence effect of the individual conditions on the decision outcome, non-trivial decisions, i.e., decisions with more than one condition are the interesting once.
 In fact, if a decision has only one condition, it is not possible to differentiate MC/DC from DC. The column \textit{Non-trivial decisions} gives the number of the decisions (out of $m$) that have at least two conditions in the model.

 \begin{table}
 	\centering
 	\caption{MC/DC coverage results for example CPN models}
 	\label{tb:results}
 	\begin{tabular}{| l | r | c | c | r | @{~~~~}r | l|}
 		\hline
 		~\thead{CPN \\Model} ~&~\thead{Executed\\ decisions}~ & \thead{~Model~\\ ~decisions ($m$)~} & \thead{~Non-trivial~ \\~decisions~} & \thead{~MC/DC~\\ ~(\%)~} & \thead{BC~~~~\\ (\%)~~~} & \thead{~State\\ space}~ \\ \hline
 		~Paxos~    &2,281,466~ & 27 & 11 & 37.03 ~& 40.74~ &~infinite~ \\ \hline
 		~MQTT~  & 3,870~ & 18 & 14 & 11.11~ &  22.22~ &~finite~  \\ \hline
 		~CPNABS~ & 3,716,896 ~& 32 & 13 & 59.37~ & 84.37~ &~infinite~\\ \hline
 		~DisCSP~  & 233,819 ~& 12 (10) & 5 & 45.45~ & 45.45~ &~finite~\\ \hline
 	\end{tabular}
 \end{table}

 Some of the models under test, for instance the MQTT-protocol, have alternative configurations contained in the model which can yield different coverage results with respect to which set of other global parameters.
 We record both MC/DC and BC as the ratio of covered decisions over the total number of decisions.
  
For the models with an infinite state space (the CPNABS and Paxos models), we aborted the state space exploration after two days at which point the number of arcs and guards expression executed no longer seemed to increase the coverage metrics.

\subsection{Discussion of Results}

\begin{figure}[tbhp]
	\centering
	\subfloat[CPN ABS model]{\includegraphics[width=.5\linewidth]{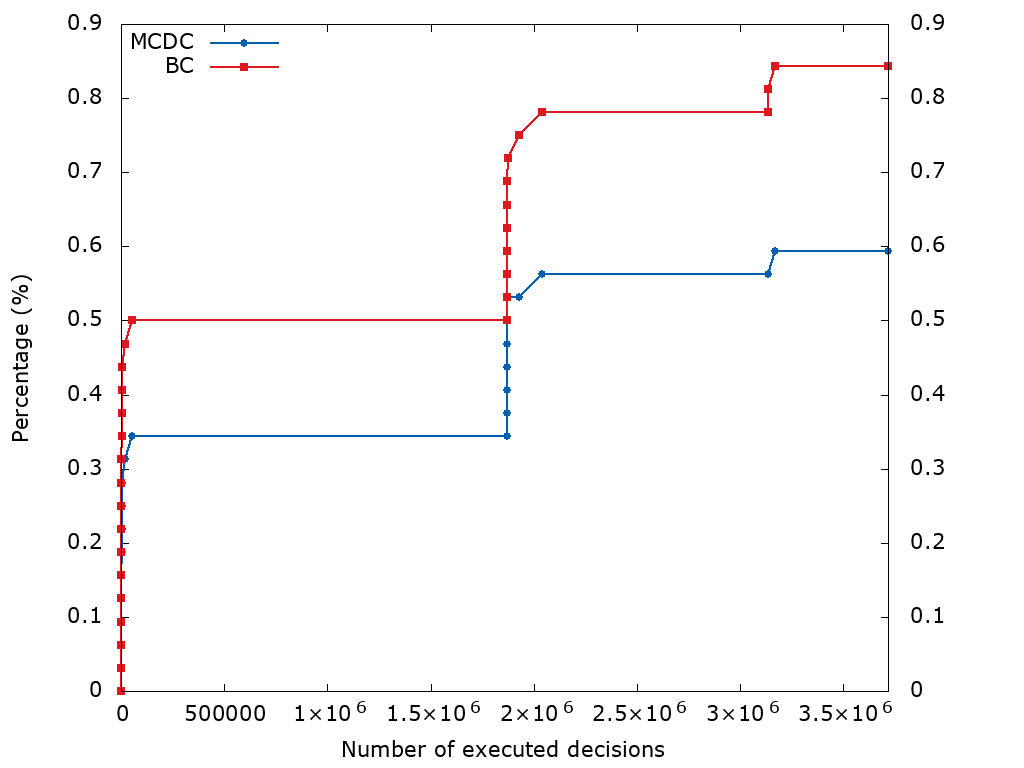}}%
	\subfloat[Paxos model]{\includegraphics[width=.5\linewidth]{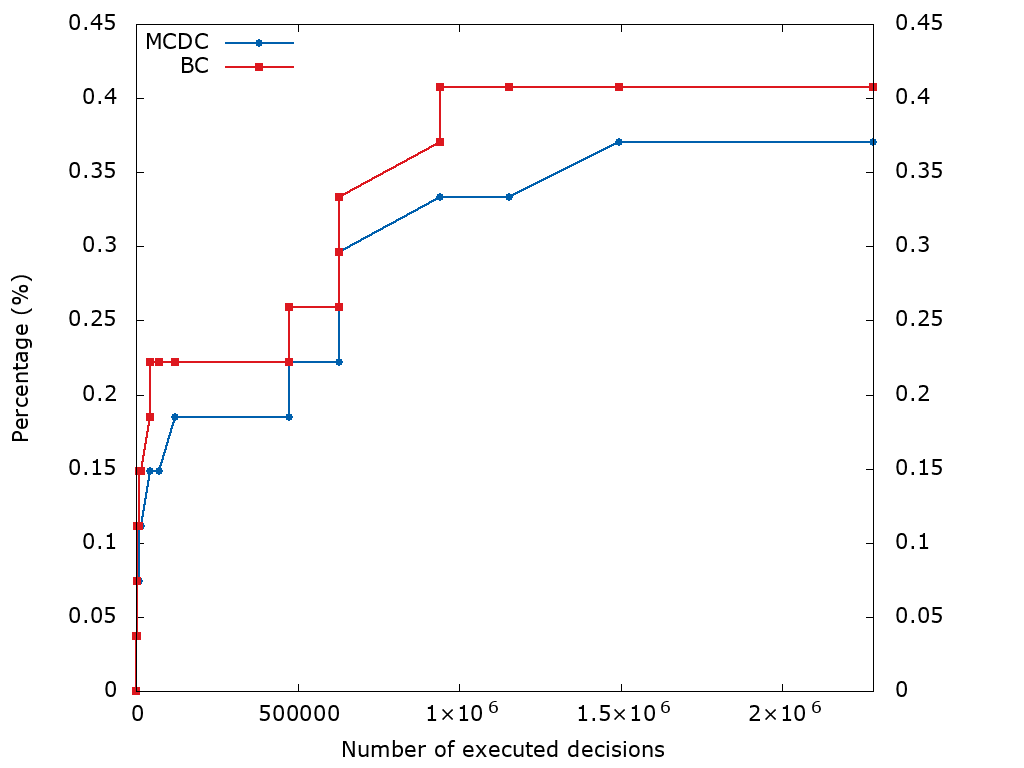}}\\%
	\subfloat[MQTT model]{\includegraphics[width=.5\linewidth]{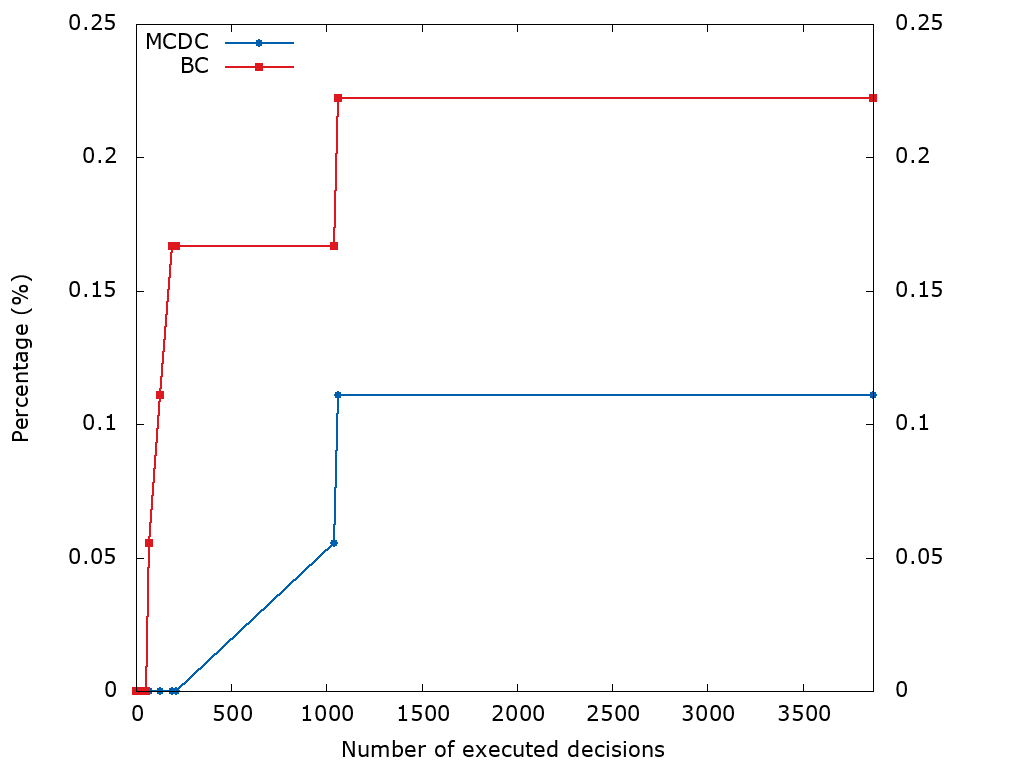}}%
	\subfloat[DisCSP WCS model]{\includegraphics[width=.5\linewidth]{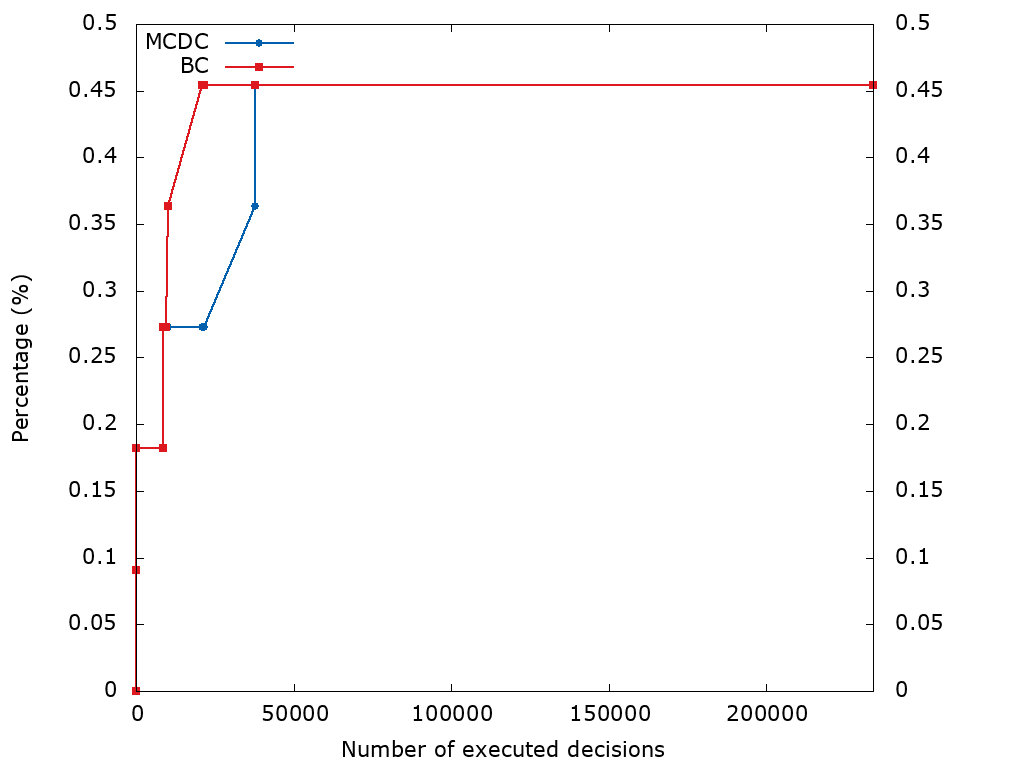}}%
	\caption{MC/DC and BC coverage versus number of executed decisions}
	\label{fig:result}
\end{figure}

MC/DC is covered if all the conditions in the decision are tested once true and once false with the independence effect on the outcome.
BC is covered if all the branches are taken at least once. This makes MC/DC a stronger coverage criterion compared to BC.
Figure \ref{fig:result} shows the graphical representation of our resulting coverage metrics as the percentage of covered decision with respect to the number of executed decisions in guard and arcs for both MC/DC and BC.
The plots show that the covered decisions increase as the model (and hence the decisions) are being executed.
Note that the x-axis does not directly represent execution time of the model:
the state space explorer already prunes states that have been already visited (which takes time), and hence as the state space exploration progress the number of expressions evaluated per time unit will decrease.
For all the models, BC is higher compared to MC/DC, which complies with their definitions and criteria.
In case the expression was executed with the same outcome, the coverage results do not increase, since those test cases have already been explored. 
For the DisCSP algorithms, we tested one of the presented models, DisCSP weak‐commitment search (WCS), since it is considered as the prime algorithm.
Two factors affect the coverage percentage results presented for these models:

\begin{enumerate}
\item The tested models had no clear test suites, they might be lacking test cases to cover the remaining conditions.
Depending on the purpose of each model, some of the test cases may not be relevant for the model or the model may not even have been intended for testing.
This could be solved by guiding the SSE for test cases generation for the uncovered decision (see discussion of future work).

\item The models might be erroneous in the sense that some parts (conditions) in the model are never or only partially executed.
For example in the DisCSP model, there are two decisions which never appear in the logs, because they were never executed, and we cannot tell if this was intentionally or not.

\end{enumerate}

A main results of our analysis of the example models is, however, that none of the models (including those for which the state space could be explored) have full MC/DC coverage. This confirms our hypothesis that code coverage of net inscriptions of CPN models adds value to the analysis. A full state space exploration represents the current limit of capabilities that CPN Tools has to exercise (test) a CPN model. Our results shows that even in presence of full state space exploration, we may still find expressions that are not MC/DC covered.  From the perspective of the present paper, it confirms the relevance and added value of performing coverage analysis net inscriptions of CPN models. A natural next step in a model development process would be to revisit the decisions that are not MC/DC covered and understand the underlying reason. Since the scope of this paper is limited to the coverage analysis, test cases case generation and fault finding in these models will be addressed in future work.

%% file: related.tex
\label{sec:related}

Coverage analysis has attracted attention in both academic and industrial research, especially the MC/DC criterion is highly recommended and commonly used in safety critical systems, including as avionic systems~\cite{pothon2012178c}.
However, there is a limited number of research addressing model-based coverage analysis.
Ghosh~\cite{GhoshSRE03} expresses test adequacy criteria in terms of model coverage and explicitly lists \emph{condition coverage} and \emph{full predicate coverage criterion} for OCL predicates on UML interaction diagrams, which are semantically related to CPNs in that they express (possible) interactions. Test cases were not automatically generated.
In \cite{10.1007/978-3-642,6915770}, the authors present an automated test
generation technique, MISTA (Model-based Integration and System Test Automation) for integrated functional and security
testing of software systems using high-level Petri nets as finite state test models.

Chilenski ~\cite{Chilenski01aninvestigation}, investigated three forms of MC/DC including Unique-Cause (UC) MC/DC, Unique-Cause + Masking MC/DC, and Masking MC/DC. Moreover, other forms of MC/DC has been discussed in ~\cite{10.1145/2554850.2555004}, where a systematic literature review on MC/DC was conducted. More than 70 papers were reviewed and 54 of them discussed MC/DC definitions and the remaining are only focusing on the use of MC/DC in fault detection. We presented in~\cite{8914094}, a tool that measures MC/DC based on traces of C programs without instrumentation. However, there is a limited number of studies on model coverage analysis based on MC/DC, and to the best of our knowledge, none has been conducted on MC/DC analysis of CPNs models.

Simulink~\cite{Mathmmodelcov,Simulmodelcov} supports recording and visualising various coverage criteria including MC/DC from simulations via the Simulink Design Verifier. It also has two options for creating test cases to account for the missing coverage in the design: either creating the test cases manually, or generating them automatically using the Simulink Design Verifier. Heimdahl and George~\cite{10.5555/1025115.1025217} performed an experiment using an example of a Flight Guidance System (FGS) mode-logic model to generate and reduce a test-suite reduction for a variety of structural coverage criteria focusing on effects on test quality and implications for testing. However, their findings had obvious threats to the external validity that prevents them from generalising their observations.
Test coverage criteria for autonomous mobile systems based on CPNs was presented by Lill et al.\ in~\cite{testcov}.
Their model-based testing approach is based on the use of CPNs to provide a compact and scalable representation of behavioural multiplicity to be covered by an appropriate selection of representative test scenarios fulfilling net-based coverage criteria. Sim{\~a}o et al.~\cite{Simo2003AFO1} provide definitions of structural coverage criteria family for CPNs, named CPN Coverage Criteria Family. These coverage criteria are based on checking if all-markings, all-transitions, all-bindings, and  all-paths are tested at least once. Our work is different from the above presented work in that we are analysing the coverage of net inscriptions (conditionals in SML decisions) in CPN models based on structure coverage criteria defined by certification standards, such as DO-178C~\cite{rierson2013developing}.


%% file: summary.tex
\label{sec:summary}

We have presented a new approach and a supporting software tool to measure MC/DC and branch coverage (BC) of SML decisions in CPN models. There are three main contribution of this paper: 1) We provide a library and annotation mechanism that intercepts evaluation of Boolean conditions in guards and arcs in SML decisions in CPN models, and record how they were evaluated; 2) we present a post-processing tool that computes the conditions truth assignment and checks whether or not particular decisions are MC/DC-covered in the considered executions of the model; 3) we collect coverage data from publicly available CPN models and report whether they are MC/DC and BC covered.

Our experimental results show that, our library and post-processing tool can find how conditions were evaluated in all the net inscriptions in CPN models and measure MC/DC and BC. Results reveal that the MC/DC coverage percentage is quiet low for all the four CPN models tested.
This is interesting because it indicates that developers may have had different goals when they designed model, and that the model only reflects a single starting configuration.
We can compare this with the coverage of regular software: running a program will yield \emph{some} coverage data, yet most programs will have to be run with many different inputs to achieve adequate coverage.

To the best of our knowledge, this is the first work on coverage analysis of CPN models based on BC and MC/DC criteria. This work highlighted that coverage analysis is interesting and useful for CPN models, not only in the context of showing the covered guard and arcs SML decisions, but also the effect of conditionals in SML decisions on the model outcome and related potential problems.
\subsection*{Outlook}
Our general approach to coverage analysis presents several directions forward which would help developers get a better understanding of their models:
firstly, while generating the full state space is certainly the preferred approach, this is not feasible if the state space is inherently infinite or too large.
Simulation of particular executions could then be guided by results from the coverage and try to achieve higher coverage in parts of the model that have not been explored yet.
However, while selecting particular transitions to follow in a simulation is straight-forward, manipulating the data space for bindings used in guards is a much harder problem and closely related to test case generation (recall the CPNs also rely on suitable initial states, which are currently given by developers).
Making use of feedback between the state of the simulation and the state of the coverage would, however, require much tighter integration of the tools. A related direction is to consider visualising coverage information in the graphical user interface:
\CPNTools{} already supports a broad palette of visual options that could be used, e.g., to indicate successful coverage of guards through colour, or the frequency that transitions have been taken through their thickness~\cite{WangiFM19}.

As for the measured coverage results, it would be interesting to discuss with the original developers of the models if the coverage is within their expectations.
While on the one hand low coverage could indicate design flaws, on the other hand our testing may not have exercised the same state space as the original developers did:
they may have used their model in various configurations, whereof the state of the \texttt{git} repository just represents a snapshot, or we did not discover all possible configurations in the model.
In the future, we may also try to generate test-cases specifically with the aim to increase coverage.
